\documentclass[titlepage]{csetr}

\usepackage{graphicx}
\usepackage{colortbl}
\usepackage{subfigure}
\usepackage{amsmath}
\usepackage{algorithm}
\usepackage{algorithmic}
\usepackage[table]{xcolor}
\definecolor{lightgray}{gray}{0.9}
\newtheorem{axiom}{Axiom}

\newtheorem{definition}{Definition}

\renewcommand{\and}{\hspace{.5cm}}

\trnumber{201204}
\trdate{February 2012}
\title{%
An Analytic Approach to People Evaluation in Crowdsourcing Systems
}
\author{%
  Mohammad Allahbakhsh$^1$ \and %
  Aleksandar Ignjatovic$^1$ \\ %
  Boualem Benatallah$^1$ \and %
  Seyed-Mehdi-Reza Beheshti$^1$\\ %
  Norman Foo$^1$\and
  Elisa Bertino$^2$\\[2em]
  $^1\, $University of New South Wales\\ Sydney 2052, Australia \\%
  \email{\{mallahbakhsh,ignjat,boualem,sbeheshti,norman\}@cse.unsw.edu.au}\\ \\
  $^2\,$Purdue University, West Lafayette, Indiana, USA \\%
  \email{bertino@cs.purdue.edu}\\[3cm]
}

\date{}
\begin{document}
\maketitle

\begin{abstract}
Worker selection is a significant and challenging issue in crowdsourcing systems. Such selection is usually based on an assessment of the reputation of the individual workers participating in such systems. However, assessing the credibility and adequacy of such calculated reputation is a real challenge. In this paper, we propose an analytic model which leverages the values of the tasks completed, the credibility of the evaluators of the results of the tasks and time of evaluation of the results of these tasks in order to calculate an accurate and credible reputation rank of participating workers and fairness rank for evaluators. The model has been implemented and experimentally validated.
\end{abstract}

\section{Introduction}
Crowdsourcing involves receiving, incorporating and consolidating contributions from a large crowd with varied levels of expertise\cite{csfirst}. Processes like building artifacts, evaluating items, and executing tasks are some instances of crowdsourcing processes\cite{cswww}. For instance, Amazon Mechanical Turk\footnote{http://www.mturk.com}(MTurk) provides on-demand access to task forces for micro-tasks such as image recognition, language translation, etc. Several organizations including DARPA and various world health and relief agencies are using platforms such as MTurk and Ushahidi\footnote{http://ushahidi.com/} to crowd-source information through multiple channels, including SMS, email, Twitter and the Web in general.

To crowdsource a task, the task owners, also called \emph{the Requesters}, submit tasks which they need done to a crowdsourcing platform. Another group of people, called \emph{workers}, contribute to solving tasks; the result of
solving a task is also called \emph{the outcome}. Requesters evaluate the outcomes and may reward workers whose outcomes have been accepted\cite{crowdforge}. The overall quality of the outcome of a crowdsourced task depends on the quality of the workers and on processes that govern the task creation, selection of workers, coordination of subtasks including reviewing intermediary outcomes, aggregation of individual contributions etc.

Reputation is a popular method for evaluation of the quality of workers in existing crowdsourcing platforms\cite{ContentDriven}. Reputation is used as an indicator of  community-wide judgment on worker performance\cite{RepSurvey}.  Existing techniques for computing workers reputation can be categorized in two categories: outcome based and user based approaches.

In outcome based approaches, worker reputation is determined based on the analysis of the outcome task performed by the worker \cite{ContentDriven,www07content}.  For example, the reputation of Wikipedia contributors is in general computed based on the quality of the content produced by the contributors \cite{ContentDriven,www07content,wikitrust}.

WikiTrust\cite{wikitrust,www07content} as an example of tools calculate reputation based on quality of the contents, employs two reputation ranks, one for users and one for content. The reputation of the user is build based on the quality of the changes she makes in the content. If the change she made is preserved by consequent editors, she will gain reputation otherwise she looses reputation. For the contents, if a content is edited by a reputable user it gains reputation and if a low reputation user edits a content, the content looses reputation.

In user based approaches, worker reputation is determined using feedbacks received explicitly or implicitly from other community members \cite{ebay, AleksRep}.

eBay reputation model\cite{ebay} is an example of using user based feedbacks to build the reputations. After any transaction in eBay, the seller and the buyer can rate each other by +1(positive), 0 (neutral), -1 (negative) along with a line of comment. Based on these evaluations, two reputations are built for every user, one as a seller and another one as a buyer. eBay also has a ranking criteria that give every seller 4 types of rank, each rank of value between 1 and 5, intended to reflect the quality of services that the seller has provided to the buyer, such as postage time or postage and handling charges. eBay uses a simple averaging model to calculate reputation of the users. There are also some reputation models which employ an adaptive averaging model for reputation calculation \cite{AleksRep}.

In general, existing reputation approaches focus on certain aspects of quality control. To obtain a fine grain view of different quality aspects, several parameters and metrics need to be considered.
In addition, existing reputation systems suffer from several problems, like lack of sufficient information, bias in user interests or evaluator dishonesty \cite{QualityinAMT, Qualityinsocualmedia,biased}. Although several solutions \cite{ebay,crowdflow,shengtrust,AleksRep,eigentrust} addressing these issues are proposed, dealing with these issues, in particular with malicious evaluators, is still a challenge.

Crowdsourcing systems usually experience malicious activities\cite{QualityinAMT, Qualityinsocualmedia}. The evaluators (either requester, reviewer or crowd voters) may cast unfair votes on the quality of the proposed outcome. Voters may support each other by casting positive votes or attack by casting negative votes regardless of the quality of the outcomes. Malicious manipulation of reputation can lead to inadequate worker selection which directly affects the quality of the obtained contributions. Also, such manipulation can harm community members, leaving them vulnerable to deceptive evaluators. For example, in online markets like Amazon Mechanical Turk loosing reputation results in decreasing chance of getting further jobs, possibly causing unfair loss of income. In both the existing crowdsourcing platforms, as well as in research prototypes, this issue is not fully addressed.

Finally, in existing crowdsourcing systems the trustworthiness of the workers are evaluated and proposed as a reputation rank. On the other hand, the evaluators are not assessed and there is no public rank to show how realistic and fair are the votes casted by and evaluator. The public ranks force people to be more realistic and responsible to keep their public profile and preserve their chance to be chosen as potential evaluators in further jobs. For example suppose that a list of reviewers are chosen to evaluate journal papers. If the impact of the reviews of a reviewer are displayed on her profile to show how fairly she has evaluated others, the reviewer will do the task with her best efforts, but if there are no consequences related to unfair behavior, the reviewer may put less effort and attention on the work.

In this paper, we address the problem of designing a comprehensive worker reputation assessment framework encompassing a set of adaptive and extensible algorithms which provides an informative and reliable evaluation of the reputation of the workers and fairness of evaluators and which adequately takes into account the trustworthiness of the person who gives the feedback, the time of evaluation and the values of the tasks completed.
Our solution builds upon results in user based reputation techniques and provides a more fine grained, more reliable and extensible reputation framework.

The need to consider time of execution of the tasks is due to the fact that the capabilities and trustworthiness of the workers may change in time. Thus, the feedback received from a trustworthy evaluator who has recently had an interaction with a worker is more dependable than a feedback received long time ago from an evaluator with a low level of trustworthiness.

The value of the task performed will be referred as \emph{credit}. The reputation built by contributing in high credit tasks is more reliable than a reputation built on processes with low credits [12].

In our setup, we employ a graph data model with workers and evaluators represented as nodes, and edges connecting each evaluator with workers which she has evaluated.

 We propose a novel model for reputation management in crowdsourcing systems.  In summary, the unique contributions of the paper are as follow:
\begin{itemize}

  \item
  We propose a credibility model to show how fair evaluators are when evaluating contributions of workers. We establish a pairwise relation between every evaluator and worker who have been in contact at least once. We call this relation \emph{degree of fairness}. We use majority consensus on the trustworthiness of the workers as an indicator for how close the evaluator's opinion is to community consensus. The bigger the difference between the community and worker's opinions is, the lower the degree of fairness of the evaluator will be.

  \item
  We propose an analytic model for calculating reputation of the workers in crowdsourcing systems. We define a pairwise trust rank between every evaluator and worker who have been in contact at least once. The trust rank is the aggregation of all evaluations received from evaluator on the worker's contributions in time. Then, we aggregate in community pairwise trust ranks from all evaluators to build a community wide reputation for the worker. The reputation is supported by a weight reflecting how reliable the calculated assessment of reputation is.
  We also leverage pairwise trust and degrees of fairness to calculate a rank called fairness rank for every evaluator to show the fairness of evaluator when assessing other community members.

  \item
  We develop an efficient and robust algorithm for computing such reputations. Our experimental results confirm that our model is robust against unfair or inaccurate evaluations, to an extent surpassing two most commonly used existing methods (eBay and Pagerank).
\end{itemize}
The remainder of the paper is organized as follow: In section 2 we formulate the problem. In section 3 we calculate the pairwise degree of fairness. Section 4 proposes calculation of pairwise trust. We propose the model for calculating reputations in section 5. Section 6 describes algorithmic model implementation and presents evaluation experiments. Related work is studied in section 7. We discuss the performance of our method and and conclude in section 8.
\section{Problem Formulation}
\subsection{Assumptions, Definitions and Setup}
In this section we articulate our assumptions and define the terms we use in this study. We also propose a summary of human intuition pertaining to trust and argue that such intuition is properly captured by our axioms. This shows rationality of trust ranks as computed using methods which satisfy the axioms we formulate.

\subsubsection{Assumptions and Definitions}
There are some terms used in this paper which may have different meanings in different contexts where they might be used.  For that reason we now specify the intended meaning of these terms in the remainder of the paper.

An \textbf{Evaluator} is the person who assesses the quality of a worker's contribution for a task. The role of an evaluator can be performed by several types of participants: by the requester who submitted the task, or by other community members such as workers or other requesters.

Such evaluation may be different from a requester to a worker, because their roles and interests give them different slants. The workers usually evaluate the intermediary results of the tasks while the decision on the quality of the final result is made by the requester. Thus, the \emph{granularity} of the work assessed by workers and by requesters is different. The other difference comes from different \emph{motivation}. Requesters and workers have different incentives for contributing to the evaluation process. Workers may participate in a crowdsourcing tasks to gain money, points or reputation but the requesters are the people who need the best final outcomes of the task at lowest cost. Even workers can be different in terms of their overall expertise and their experience in the system. The impact of these differences will be taken into account in a future refinement of our work;  since in this paper we want to introduce the fundamentals of our methodology, we will simplify our framework by not distinguishing systematic differences in evaluations provided by workers and requesters; however, the feasibility of such refinements will be quite transparent.

\textbf{Unfair evaluation} in crowdsourcing systems is an evaluation which does not adequately reflect the quality of the evaluated outcome, and it usually comes from malicious alter motives. Its purposeful form is called an \emph{attack}. An example is when users who have lost their reputation due to deceptive behavior attempt to boost their reputation by a Sybil attack. In such an attack they may create a requester account and submit some tasks, do them as a worker and accept them as a requester thus boosting their reputation\footnote{http://www.behind-the-enemy-lines.com/2010/10/be-top-mechanical-turk-worker-you-need.html}. Unfair evaluation might be motivated by various reasons. For example, workers may try to harm or support other workers or some particular requesters. Requesters, on the other hand, may evaluate contributions unfairly in order to avoid paying a fair fee for the task done, or even to harm a particular worker. While we do not study the inherent features of such attacks, we do experimentally demonstrate robustness of our method in such attacks. Such robustness is due to the inherent adaptive features of our aggregation methods.

\subsubsection{Setup}
\begin{figure*}
\centering
\includegraphics[scale=0.41]{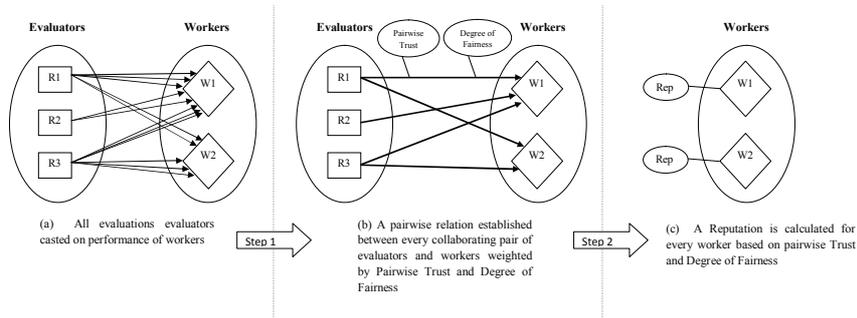}
\caption[My Image] {The process of calculating reputations.}
\label{gen}
\end{figure*}

Let us assume that a crowdsourcing system has N members. These members may submit tasks, do tasks submitted by others or evaluate others' contributions. Suppose that $N_W$ members have contributed to at least one crowdsourcing task. We call this group of members \emph{Workers} and denote by $W = \{w_j: 0 \leq j \leq N_W \}$. We also assume that $N_R$ members have evaluated at least one contribution in the system. We call this group, \emph{Evaluators} and denote them by $R = \{r_i: 0 \leq i \leq N_R \}$. Suppose that a worker $w_j$ has prepared a contribution and submitted it to the system. The $r_{i}$, an evaluator, assesses the results received form $w_{j}$ at time stamp $k$ and sends this evaluation to a Reputation Management Service (RMS); these evaluations will be taken into account for determining reputation of $w_j$. We denote this evaluation by $e_{ij}(k)$ . We assume that the $e_{ij}(k)$ is a real number in a fixed range $[0,M]$, ($ M > 0$). If $e_{ij}(k)=M$, this means that $r_i$ has determined that the quality of performance of $w_j$ warrants full trust at time $k$ for the specified job, while $e_{ij}(k)=0$ means that the quality of performance of $w_j$ warrants a complete distrust. We emphasize that $R$ and $W$ are not necessarily disjoint sets i.e. there might be some workers who have evaluated other's contributions as well as some evaluators who have contributed in crowdsourcing tasks. We just draw these sets on figure \ref{gen} as disjoint to simplify visual representation of the model for easier understanding.

To introduce our axioms, calculations and definitions, we identify the sequence of all evaluations by $r_i$ of $w_j$ at all time instances $k$ by a partial function $\vec{s}_{ij} = \{(k,e_{ij}(k)) : k\in D(\vec{s}_{ij})\}$. The domain $D(\vec{s}_{ij})$ is the collection of all time instances that contributions of W$_{j}$ has been evaluated by R$_{i}$. For the simplicity, if indices $i$ and $j$  are clear from the context, we drop them from the definition of the function and say that for every $k\in D( \vec{s} ), \vec{s}[k] = e(k)$.

Our strategy is to calculate the reputation of the workers in two steps, as shown in figure \ref{gen}. In the first step we calculate the trust of $r_i\in R$ in $w_j\in W$ for all such pairs, as well as the degree of fairness of such trust. In the second step we aggregate such pairwise trust assessments into reputation ranks of evaluated workers.

We now formulate our axioms which capture intuition on trust; numerical value of trust assessment we call the (\emph{trust rank}) and denote it by $T$.
\begin{axiom}
     \label{ax:first}
     \textbf{Strict Monotony:} If $\vec{s_1}$ and $\vec{s_2}$ are two evaluation sequences with a domain $D$, and if $\vec{s_{1}}[k]\geq \vec{s_{2}}[k]$ for all $k \in D$, then $T(\vec{s_1}) \geq T(\vec{s_2})$. In addition, if $\vec{s_1}[k] \neq \vec{s_2}[k]$, then $T(\vec{s_1}) \neq T(\vec{s_2})$.
\end{axiom}
This axiom says that better evaluations should result in higher trust ranks. For instance suppose that contributions of $w_1$ has been evaluated by two evaluators $r_1$ and $r_2$ in same time instances and the evaluation results she has received from $r_1$ always have been higher than results received form $r_2$. This axiom states that the calculated trust rank for the relation between $r_1$ and $w_1$ should be higher than trust rank of the relation between $r_2$ and $w_1$.
\begin{axiom}
     \label{ax:second}
     \textbf{Averaging:} If $\vec{s}$ is an evaluation sequence with a domain $D$, then $min(\vec{s}) \leq T(\vec{s}) \leq max(\vec{s})$.
\end{axiom}
This axiom means that the calculated trust rank should be a form of average of all evaluations included in the calculation of the trust rank. As with the usual notions of average of a set of numbers, the trust rank as the aggregation of evaluations in the time should always fall between the minimum and maximum values received as evaluations for the relation between the evaluator and the worker.
\begin{axiom}
     \label{ax:third}
     \textbf{Time Discounting:} Let $e$ and $E$ be two positive numbers and $0 \leq e < E \leq M$. Also suppose that $\vec{s_1}$ and $\vec{s_2}$ are two evaluation sequences such that $D(\vec{s_1}) = D(\vec{s_2})) = \{1,2\}$, defined as $\vec{s_1}[1] = e, \vec{s_1}[2] = E$ and $\vec{s_2}[1] = E, \vec{s_2}[2] = e$. Then $T(\vec{s1}) \geq T(\vec{s2})$.
\end{axiom}
Axiom \ref{ax:third} formalizes our intuition that older evaluations can not be more important than the recent ones. This axiom states that even when two sequences contain same evaluation values, the trust rank calculated based on these sequences are not necessarily same and it depends on the time stamp of every evaluation. In other words, from two evaluations with equal values, the recent one should have a higher weight than the old one in the calculated trust rank.

The time stamp of evaluations in $\vec{s}$ may be too large, particularly when the crowdsourcing system has been working for several years. To simplify including the time stamp of evaluations into calculations, instead of using the time stamp itself as in \cite{AleksRep}, we divide the system life time to some smaller time intervals like days, weeks, months, quarters, etc and assign an incremental index to every time interval starting from one. Then we use the index of the time interval in which the evaluations happened as the time label of the evaluation and denote it by $\vartheta_k$. The $\vartheta_k$ is smaller than $k$ and decreases complexities of including time in calculations.

We also define a constant called $q$, ($q \geq 1$); the value assigned to $q$ determines how fast the importance of a past evaluations decreases as the time progresses, and is system dependant. Suppose that the ``half life'' of the importance of evaluations in a system is $t$ i.e. the importance of an evaluations after $t$ intervals decreases to a half of its original value. The constant $q$ is then calculated using the following equation:
\begin{equation}
\label{eqq}
q = 2^{1/t}
\end{equation}
The reason will be clear from equation \eqref{eq1}.

\subsection{Example Scenario}
\label{scenario}
Voting is one of the popular crowdsourcing tasks\cite{votingcssurvey}. In this kind of tasks, the opinions of the crowd are collected to help requesters making better decisions. In Wikipedia\footnote{http://wikipedia.org}, the voting process is used to elect administrators. Every registered user can nominate herself or another user for being an administrator in Wikipedia and initiate and election. The other users participate in the election and cast their votes on the eligibility of nominee to be an administrator in the Wikipedia and if the majority of the users recognize her eligible, she will become a Wikipedia administrator. In this crowdsourcing task, the requester is the nominator, the worker is the nominee, evaluators are are voters, the task is evaluating the eligibility of the nominee for adminship in Wikipedia and contribution is the nominee's request. For instance, `andyl', a Wikipedia user nominates `cjcurrie' another user for adminship. 27 users take part in the election and cast their votes. Finally, `cjcurrie' becomes an administrator in Wikipedia as the result of receiving 22 positive, one negative and four neutral votes (see table \ref{log}).

We use the interaction log of Wikipedia Adminship Election data\footnote{http://snap.stanford.edu/data/wiki-Elec.html} that are collected by Leskovec et. al for behavior prediction in online social networks\cite{datasetcollector}, referred in the following as WIKILog. WIKILog contains about 2,800 elections with around 100,000 total votes and about 7,000 users participating in the elections either as voter or nominee. We will use the WIKILog in the paper to demonstrate how it is possible to use the proposed framework to calculate people reputations in crowdsourcing systems. For example, we will show how it is possible to: (i)~calculate pairwise trust between evaluators and workers; (ii)~calculate the degree of fairness between evaluators and workers; and (iii)~calculate reputation of the workers. Table \ref{log} shows an example of WIKILog data.

\begin{table*}
  \begin{center}
    \rowcolors{1}{}{lightgray}
    \begin{tabular}{p{1.2cm}  p{1.0cm}  p{1.0cm}  p{1cm}  p{1.0cm}  p{1.2cm}  p{0.7cm}  p{1.2cm}  }
    \hline
    Election \newline Closing Time&Nomi-nator&Nominee&Election Status&Voter  ID&Voter Name&Vote&Time of Voting \\
    \hline
    2004-09-21 01:15:53 & andyl & cjcurrie & 1 & 3 & ludraman & 1 & 2004-09-14 16:26:00 \\
    2004-09-21 01:15:53 & andyl & cjcurrie & 1 & 25 &  blankfaze& -1 & 2004-09-14 16:53:00 \\
    ....&....&....&....&....&....&....&.... \\
    2005-07-05 00:11:04 & lst27 & lst27& 0 & 33 & chmod007 & 1 & 2004-09-14 22:12:00 \\
    2005-07-05 00:11:04 & lst27 & lst27& 0 & 82 & xiaopo & 0 & 2004-09-16 17:34:00 \\
    ....&....&....&....&....&....&....&.... \\
    \hline
    \end{tabular}
  \end{center}
  \caption{Example of Wikipedia Adminship Election Log (WIKILog).}
  \label{log}
\end{table*}

\subsection{Data Model}
We model crowdsourcing entities (mainly evaluators and workers) in a crowdsourcing process log and their relationship as a directed graph $G =(V, E)$ where $V$ is a set of nodes representing entities and $E$ is a set of directed edges representing relationships between nodes. Following we describe how to: (i)~extract (evaluator and worker) nodes from the crowdsourcing log; (ii)~establish relationship between generated nodes; and (ii)~calculate the reputation of the people. These two steps are illustrated in figure \ref{gen}.

\textbf{Step~1: Preprocessing.} The aim of preprocessing of the crowdsourcing log is to generate a graph by considering the set of workers and evaluators in the log as nodes of the graph, and the presence of certain relationships between them encoded as edges between nodes, such as $r_i$ is an evaluator of $w_j$, or $r_i$ trust $w_j$. In order to preprocess the WIKILog, we performed the following two steps: (i) we generate graph nodes (i.e., evaluators and workers)by extracting interactions and their attributes from the log and formed the set of graph nodes (vertices), one for each person (but with no relations between nodes); and (ii)~we generate relationship between nodes: we used our previous work~\cite{BPM11,Motahari:ProcessSpace2}, to formulate the relationships between any pairs of nodes in the crowdsourcing log. In particular, we used \emph{correlation condition}~~\cite{Motahari:ProcessSpace2} as a binary predicate defined on the attributes of graph nodes (evaluators and workers) that allows to identify whether two or more nodes are potentially related.

Then we used the technique proposed in~\cite{BPM11} to establish a pairwise relationship between every evaluator $r_{i}$ and worker $w_{j}$ whose contribution has been evaluated by $r_{i}$ at least once during system life time. This relation is weighted by two attributes. The first attribute is a pairwise trust $\tau_{ij}$ which shows in what extent the $r_i$ trusts in the $w_j$. We call this trustworthiness degree \emph{Trust Rank}. The second attribute is the degree of fairness $\varphi_{ij}$ which shows how fair the $r_i$ has been when evaluating the $w_j$.

\textbf{Step~2: Calculating Reputation Degree.} In the second step, using the pairwise trusts and also degree of fairness values calculated for every pairwise relation between evaluators and workers up to time $n$ , we calculate a reputation degree $\rho$(n) for every worker; here $n$ is the time at which the reputation is calculated e.g. current time stamp of the system. Every worker's reputation is supported by an assigned weight $\omega_j$ showing how many evidences and trust ranks have been considered to build that reputation.

We also calculate a Fairness rank $\gamma_i(n)$ for every evaluator to show how fair she has been when evaluating others. The fairness rank is supported by a weight called weight of fairness and denoted by $\psi_i(n)$ showing how dependable is the calculated fairness rank.

\section{Pairwise Degree of Fairness}
The RMSs must be robust against unfair evaluations and while helping requesters find high quality workers, an RMS must protect workers against unfair evaluators as well. The proposed credibility model in this section addresses this problems.

Research shows that in a normal community the majority consensus usually is dependable and is close enough to experts' judgments about what is good and what is bad\cite{MajorityConsensus,shengtrust}. Applied to reputation, majority of the evaluators provide a realistic and dependable evaluation of the performance of a worker. So, we use the majority consensus as a measure for checking the credibility of the evaluations provided between pairs of evaluators and workers.

We calculate a \emph{degree of fairness} for every pairwise relation between evaluators and workers. For example if an evaluator has assessed contributions of $n$ workers, we will create $n$ degree of fairness relations one for every established relation. The degree of fairness between $r_i$  and $w_j$ is denoted by $\varphi_{ij}$. Degree of fairness shows how credible is the trust rank calculated between an evaluator and a worker. We will use $\varphi_{ij}$ to show how accurate are the trust feedbacks that $r_i$ has given to $w_j$.

We calculate degree of fairness in four steps. At first, we calculate the average of all evaluations given to a particular worker, say $w_j$. The equation \ref{eq3} shows how the average is calculated. The $D_j$ is the set of all evaluators have assessed contributions of $w_j$.
\begin{equation}\label{eq3}
\bar{e_j} = \frac{\sum_{l \in D_j} e_{lj}(k)}{|D_j|}
\end{equation}
In the second step, we calculate the average of all evaluations given to $w_j$ by $r_i$ using the equation \ref{eq4}. $N_{ij}$ is the number of evaluations that $r_i$ has done on the trustworthiness of the $w_j$.
\begin{equation}\label{eq4}
\bar{e_{ij}} = \frac{1}{N_{ij}} \sum_{}{e_{ij}(k)}
\end{equation}
In the third step, we calculate the standard deviation of all evaluations given to a worker as we show in equation \ref{eq5}
\begin{equation}\label{eq5}
SD_j =\sqrt{ \frac{\sum_{k \in D_j} (\bar{e_{kj}} - \bar{e_{j}}) ^2 }{|D_j|}}
\end{equation}
Finally, we define and calculate degree of fairness of relations between evaluators and workers regarding the Equations \ref{eq3}, \ref{eq4} and \ref{eq5}.
\begin{definition}
    \label{df:crd}
    Suppose that the $r_i$ has assessed contributions of $w_j$ in the time. The \emph{Degree of Fairness} between $r_i$ and $w_j$ shows how fairly $r_i$ has evaluated the contributions of $w_j$ and is denoted by $\varphi_{ij}$. The $\varphi_{ij}$ is calculated as follow:
    \begin{equation}\label{eq:crd}
      \varphi_{ij} = \left\{
      \begin{array}{l l}
        \frac{\bar{e_j} - SD_j - \bar{e_{ij}}}{M} & \quad \text{if $\bar{e_{ij}} < (\bar{e_j} - SD_j)$}\\
        1 & \quad \text{if $(\bar{e_j} - SD_j) \leq \bar{e_{ij}} \leq (\bar{e_j} + SD_j)$}\\
        \frac{\bar{e_{ij}}- (\bar{e_j} + SD_j)}{M} & \quad \text{if $ (\bar{e_j} + SD_j) < \bar{e_{ij}} $}\\
      \end{array} \right.
    \end{equation}
\end{definition}
According to Equation \ref{eq:crd}, the averages fall in $\bar{e_j} \pm SD_j$ are trustworthy and dependable but the averages that fall out of that range are of very low credibility and their impact on reputation are decreased dramatically. The $\varphi_{ij}$ shows how close to the majority consensus is the judgment of $r_i$  about $w_j$'s trustworthiness. We use degree of fairness to reduce the effect of evaluations generated by evaluators who do not agree with the majority consensus.
\section{Pairwise Trust}
Regarding the analytic model proposed in \cite{AleksRep}, axioms stated above, defined constant $q$ and also time label $\vartheta_k$, we define the trust rank between evaluators and workers as follows:
\begin{definition}
     \label{df:first}
     Suppose that $\vec{s}$ is the sequence of all evaluations from $r_i$ on performance of $w_j$ at time instances from the domain $D$. We define \emph{Trust Rank} $\tau_{ij} = T(\vec{s})$, intended to be a measure how much $r_i$ trusts $w_j$, as follows:
    \begin{equation}\label{eq1}
    \tau_{ij} = \frac{\sum_{k \in D} e_{ij}(k) q^{\vartheta_k}}{\sum_{l \in D} q^{\vartheta_l}}
    \end{equation}
\end{definition}
Pairwise trust $ \tau_{ij}$ is the aggregation of all evaluations received from $r_i$ on the contributions of $w_j$. The aggregation formula given by equation \ref{df:first} satisfies all three axioms we propose (see \cite{AleksRep} for a proof).

\begin{definition}
     \label{df:wotr}
     Suppose that $\vec{s}(n)$ is the sequence of all evaluations from $r_i$ on performance of $w_j$ up to the time instance $n$. We define \emph{Weight of Trust Rank}, intended to be a measure how dependable the calculated trust rank $\tau_{ij}$ is, based on the evaluations involved in calculation of the trust rank, and denote it by $\omega_{ij}(n)$. The value $\omega_{ij}(n)$ is calculated as follow:
    \begin{equation}\label{eq:wotr}
    \omega_{ij}(n) = \frac{ \sum_{l \in D} q^{\vartheta_l} h(c(i,j,l))}{q^{\vartheta_n}}
    \end{equation}
\end{definition}

The $c(i,j,l)$ denotes the credit related to the task has been done by $w_j$ and evaluated by $r_i$ at time $l$. Function $h$ is a strictly increasing function that defines how the $c(i,j,l)$ must be considered in the weight of the trust rank; in our experiments $h(x)=x$, i.e., $h(c(i,j,l)) = c(i,j,l)$.

Equation \ref{eq:wotr} counts the number of evaluations that have been used for building trust rank, taking into account their time label. Since we believe that recent evaluations should have more importance in the system than the older ones, we accumulate the time weighted number of evaluation rather than simply counting them.

The other parameter that is important in the weight of the trust rank is the amount of credit (e.g. money) that has been paid for doing the crowdsourcing task. Since $h(x)$ is increasing, more credit will lead to more influence on the resulting weight of the trust rank.

\section{Reputation of the Workers}
In this section we provide an explicit calculation for obtaining reputation for every worker. The reputation of a worker is an aggregation of all trust ranks between the worker and all evaluators with whom she is related, weighted appropriately by the corresponding weight of trust.
\begin{algorithm}[t]
\caption{ Worker's Reputation Calculation}
\label{alg:wrep}
\textbf{Input:} T as the set of all pairwise trusts $\tau_{ij}$, $\omega$ as the set of the weight of all pairwise trust ranks $\omega_{ij}$, $W$ as the set of all workers and $\Phi$ as the set of all pairwise degrees of fairness of each evaluator and  each worker who was evaluated by this evaluator at least once.\\  \\
\textbf{Output:} $P$ as the set of all reputation scores ($\rho_j$) and $\Omega$ as their corresponding weights ($\omega_j$)
\begin{algorithmic}
\FORALL{$w \in W$}
\STATE $r = 0$
\STATE $w = 0$
\STATE $Tr \leftarrow$ All Trust ranks on Worker w ($T_w \subset T$)
\STATE $Wt \leftarrow$ Weight of all trust ranks in $Tr_w$ ($Wt \subset \Omega$)
\STATE $Cr \leftarrow$ Credibility of of all trust ranks in $Tr_w$ ($Cr \subset \Phi$)
\FORALL{$\tau \in Tr_w$}
\STATE wupdate = $Wt_{\tau} * Cr_{\tau}$
\STATE $ r = r + \tau*wupdate $
\STATE $ w = w + wupdate $
\ENDFOR
\STATE $P_w = r/w $
\STATE $\Omega_w = w $
\ENDFOR
\RETURN $P$ and $\Omega$
\end{algorithmic}
\end{algorithm}

Since the reputation is a community wide judgement of trustworthiness of a worker, we calculate the reputation of workers by aggregating all trust ranks between the worker and all evaluators who has evaluated her.
\begin{definition}
     \label{df:rep}
     Suppose that $D_j$ is the set of all evaluators who have assessed contributions of $w_j$. We define the \emph{Reputation} $\rho_{j}(n)$ as the community wide judgement of trustworthiness of $w_j$ up to time instance $n$ and calculate it as follow:
    \begin{equation}\label{eq:rep}
    \rho_{j}(n) = \sum_{i \in D_j}{\frac{\omega_{ij}(n) \varphi_{ij}}{\sum_{l \in D_j}{\omega_{lj}(n) \varphi_{lj}}}\tau_{ij}}
    \end{equation}
\end{definition}

Equation \ref{eq:rep} and Algorithm \ref{alg:wrep} show that the reputation is an aggregate of the trust ranks $\tau_{ij}$  that a worker received from all evaluators, prorated by their corresponding degree of fairness, which is reflected in the value of the corresponding multiplier ${\omega_{ij}(n) \varphi_{ij}}/{\sum_{l \in D_j}\omega_{lj}(n) \varphi_{lj}}$. Such multiplier takes into account the weight of the trust rank $w_{ij}$ and the corresponding degree of fairness $\varphi_{ij}$; the denominator $\sum_{l \in D_j}\omega_{lj}(n) \varphi_{lj}$ re-normalizes the sum, making $\rho_j(n)$ a weighted average of individual trust ranks. This method of calculating trust ranks reflects our intuition that different evaluators have different credibility levels which should reflect their overall behavior in the system.

As we stated before, our model distinguishes between the reputations that are build based on high number of evaluations received from fair evaluators and reputations build based on few number of evaluations received from unfair evaluators. This is possible by providing a corresponding weight rank of every calculated reputation.
\begin{definition}
     \label{df:wor}
     Suppose that $D_j$ is the set of all evaluators who have assessed contributions of $w_j$. We define the \emph{Weight of Reputation} and denote it by $\Omega_j(n)$ to show how dependable is the calculated reputation for $w_j$ which is calculated up to time instance $n$. We consider the weight and credibility of trust ranks involved in calculating the reputation to compute its weight. The $\Omega_j(n)$ is calculate as follows:
    \begin{equation}\label{eq:wor}
    \Omega_j(n) = \sum_{l \in D_j}{\omega_{lj}(n) \varphi_{lj}}
    \end{equation}
\end{definition}
Equation \ref{eq:wor} shows that the weight of reputation is calculated by aggregating weight of trust ranks received from all involved evaluators weighted by the pairwise degree of fairness between every evaluator and the worker.

\section{Fairness of the Evaluators}
\begin{algorithm}[t]
\caption{Evaluator's Fairness Calculation}
\label{alg:rrep}
\textbf{Input:} T as the set of all pairwise trusts $\tau_{ij}$, $\omega$ as the set of the weight of all pairwise trust ranks $\omega_{ij}$, $R$ as the set of all evaluatros and $\Phi$ as the set of all pairwise degrees of fairness between evaluators and workers  \\  \\
\textbf{Output:} $\Gamma $ as the set of all fairness ranks ($\gamma _i$) and $\Psi$ as their corresponding weights ($ \psi_i $)
\begin{algorithmic}
\FORALL{$r \in R$}
\STATE $re = 0$
\STATE $wt = 0$
\STATE $Tr \leftarrow$ All Trust ranks given by $r$ ($Tr \subset T$)
\STATE $Wt \leftarrow$ Weight of all trust ranks in $Tr$ ($Wt \subset \Omega$)
\STATE $Cr \leftarrow$ Degrees of fairness of of all pairwise relations in $Tr$ ($Cr \subset \Phi$)
\FORALL{$\tau \in Tr$}
\STATE $ re = re + Wt_{\tau} * Cr_{\tau} $
\STATE $ wt = wt + Wt_{\tau} $
\ENDFOR
\STATE $\Gamma_r = re/wt $
\STATE $\Psi_r = wt $
\ENDFOR
\RETURN $\Gamma$ and $\Psi$
\end{algorithmic}
\end{algorithm}

One of the common problems in existing CS platforms is lack of evaluator assessment. Evaluators evaluate workers and try to choose workers that generate high quality contributions but there are no facilities to show how trustworthy and fair are the evaluators.  Sometimes deceptive evaluators evaluate workers unfairly and try to cheat and harm them even when they have provided high quality contributions.

An effective way to prevent evaluators from being unfair is to provide a community-wide degree of fairness for every evaluator to show how fair the evaluator is. We calculate the fairness of evaluators based on the pairwise degrees of fairness between the evaluators and workers.
\begin{definition}
     \label{df:repr}
     Suppose that $S_i$ is the set of all workers who have been involved in at least one task with the evaluator $R_i$ up to time instance $n$. We define the \emph{Fairness of Evaluator} as the community wide aggregation of the degree of fairness of the relations of the evaluator with the workers to show how fair the evaluator has been or is expected to be while working with the workers and denote it by $\gamma_i(n)$. The $\gamma_i(n)$  is calculated as follow:
    \begin{equation}\label{eq:repr}
    \gamma_i(n) = \frac {\sum_{j \in S_i}{\omega_{ij}(n) \varphi_{ij}}}{\sum_{l \in S_i}{\omega_{il}(n)}}
    \end{equation}
\end{definition}

To show the community how dependable is the calculated fairness, we provide a weight along with every fairness rank calculated for the evaluator.
\begin{definition}
     \label{df:worepr}
     Suppose that $S_i$ is the set of all workers who have been involved in at least one task with the evaluator $R_i$ up to time instance $n$. We define the \emph{Weight of Fairness} to show how dependable the fairness rank calculated for the $R_i$ is and denote it by $\psi_i(n)$. The $\psi_i(n)$ is calculated as follow:
    \begin{equation}\label{eq:worepr}
    \psi_i(n) = \sum_{l \in S_i}{\omega_{il}(n)}
    \end{equation}
\end{definition}
The $\psi_i(n)$ is built out of the weights of all pairwise degree of fairness relations between the evaluator and the workers. The weight of fairness in fact counts the number of workers that has been involved in crowdsourcing tasks with the $R_i$ weighted by the weight of pairwise trust rank between them. Algorithm \ref{alg:rrep} shows how the fairness of the requester and its corresponding weight are calculated.

\section{Implementation and Evaluation}
\begin{figure}
\centering
\subfigure[Percentages]{
\includegraphics[scale=0.45]{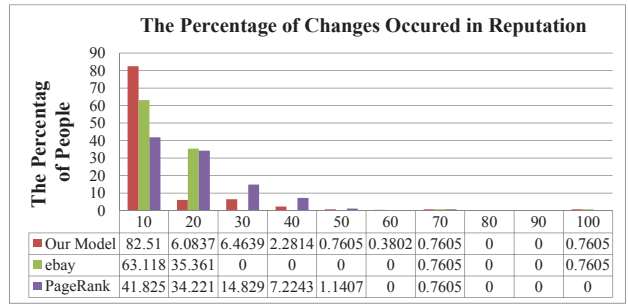}
\label{difs}
}
\subfigure[Our Model]{
\includegraphics[scale=0.40]{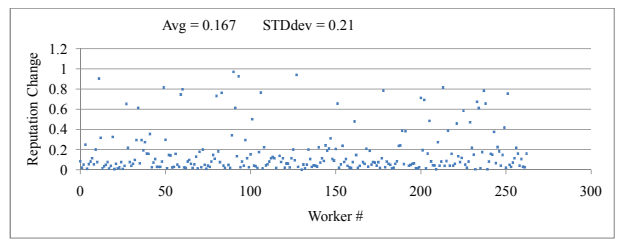}
\label{fig:om}
}
\subfigure[eBay (Normal Averaging Model)]{
\includegraphics[scale=0.40]{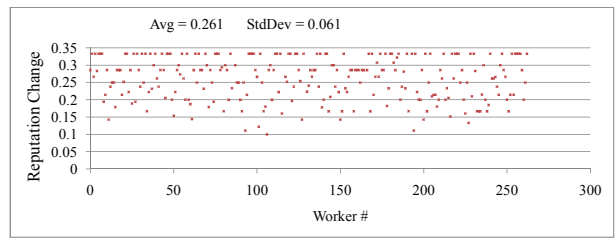}
\label{fig:ebay}
}
\subfigure[Pagerank (Adaptive Averaging Model)]{
\includegraphics[scale=0.40]{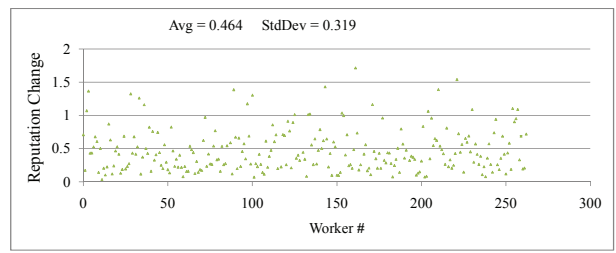}
\label{fig:pr}
}
\label{fig:difs}
\caption[]{Changes happened in reputations in three models}
\end{figure}

\subsection{Implementation}
We implemented our model using Java as the programming language and IBM DB2 as the database manager. We have used the WIKILog introduced in Section \ref{scenario} as the dataset for model evaluation. To make the data match the model we suppose that all tasks have the same credit equal to one. All the elections have happened between 2004 and 2008. To define $\vartheta_k$, we have divided the time period to half year time intervals. So, the evaluations have the time labels from one to eight. We also supposed that half life of the importance of evaluations is one year, so we have chosen the half life $t=2$, which, by equation \eqref{eqq} gives $q=\sqrt{2}$.

To evaluate our model, we compare it with two popular reputation calculation models. The first one is normal averaging model which is widely used in existing CS systems. The models used in Amazon\footnote{http://www.amazon.com}, ebay\textsuperscript{\textregistered}\footnote{http://www.ebay.com}, and lots of other online communities or markets use normal averaging. In normal averaging model reputation is the average of all votes casted on the quality of he worker's contributions. The second model is adaptive averaging model in which the votes casted by people are weighted by their reputation and then employed to calculate the reputation of the worker. The Google Pagerank model\cite{pagerank}, EigenTrust \cite{eigentrust} and our previous work \cite{AleksRep} are samples of adaptive averaging model. We have chosen Pagerank as the base of all these reputation models to compare with our model.

\subsection{Experimentation and Evaluation}
In order to evaluate the model robustness against unfair evaluations, in the first step we applied all three models to WIKILog and calculated a reputation rank for every worker in every model. Then, we added some noises to the dataset to check robustness of the models against unfair evaluations.

To add noises we added 20 percent unfair votes on all workers. To check the credibility of models we tried to manipulate reputation of people by supporting all untrustworthy people (workers with normal average reputation less than 2) by adding votes with value of 3. We also attacked all trustworthy workers (workers with normal average reputation greater or equal than 2) by adding votes with value of 1. Then we calculated the reputations again and calculated the percentage of changes happened in the reputation of the workers. figure \ref{difs} shows how the reputation of the workers have changed after adding noises to the dataset. The horizontal axis of the chart is the percentage of changes happened in the reputation of the workers and the vertical axis is the percentage of the workers who has that amount of the changes in their reputation.
\begin{figure}
\centering
\subfigure[Our Model]{
\includegraphics[scale=0.60]{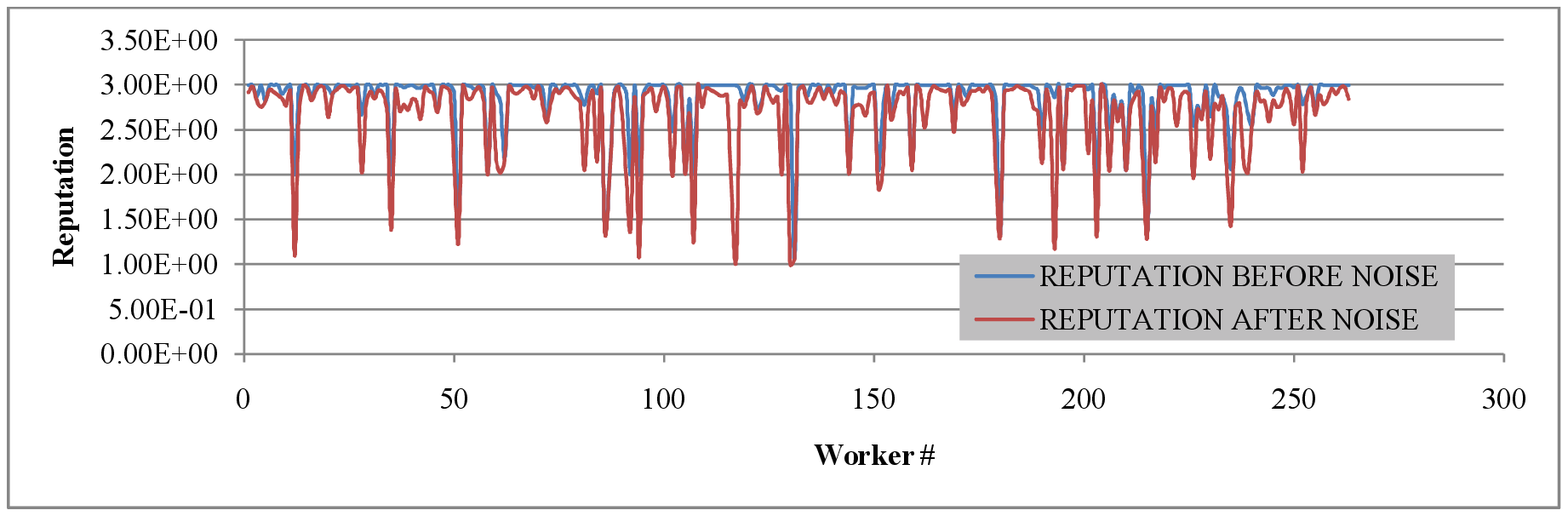}
\label{fig:omd}
}
\subfigure[eBay (Normal Averaging Model)]{
\includegraphics[scale=0.60]{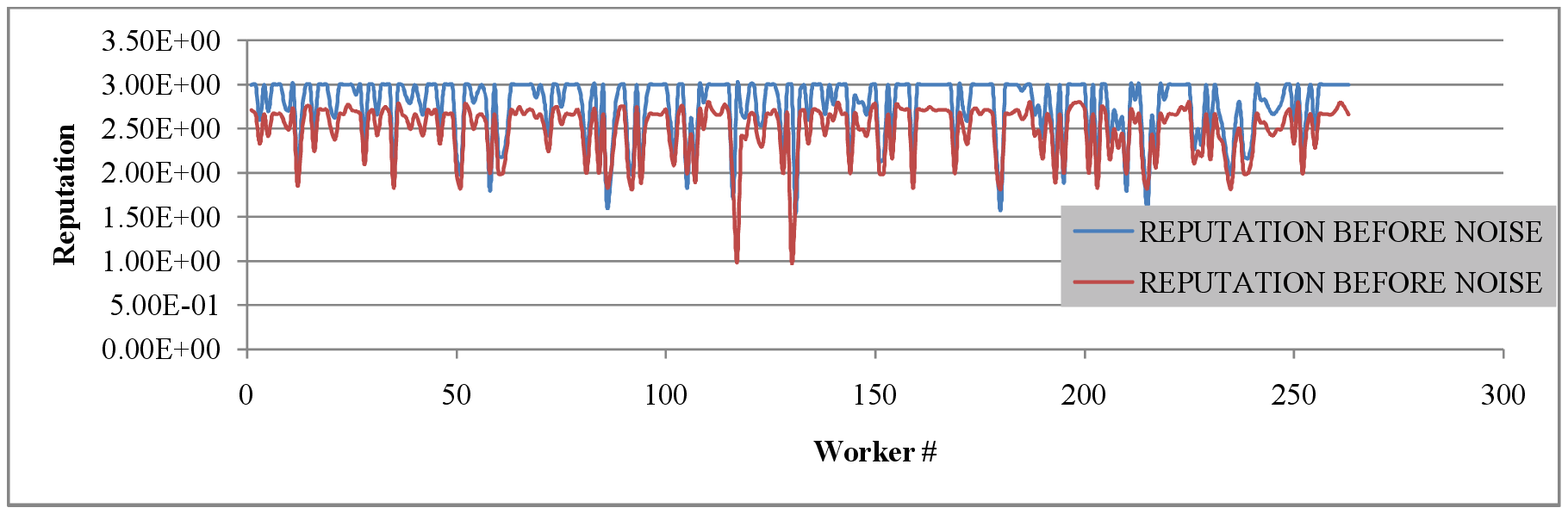}
\label{fig:ebayd}
}
\subfigure[Pagerank (Adaptive Averaging Model)]{
\includegraphics[scale=0.60]{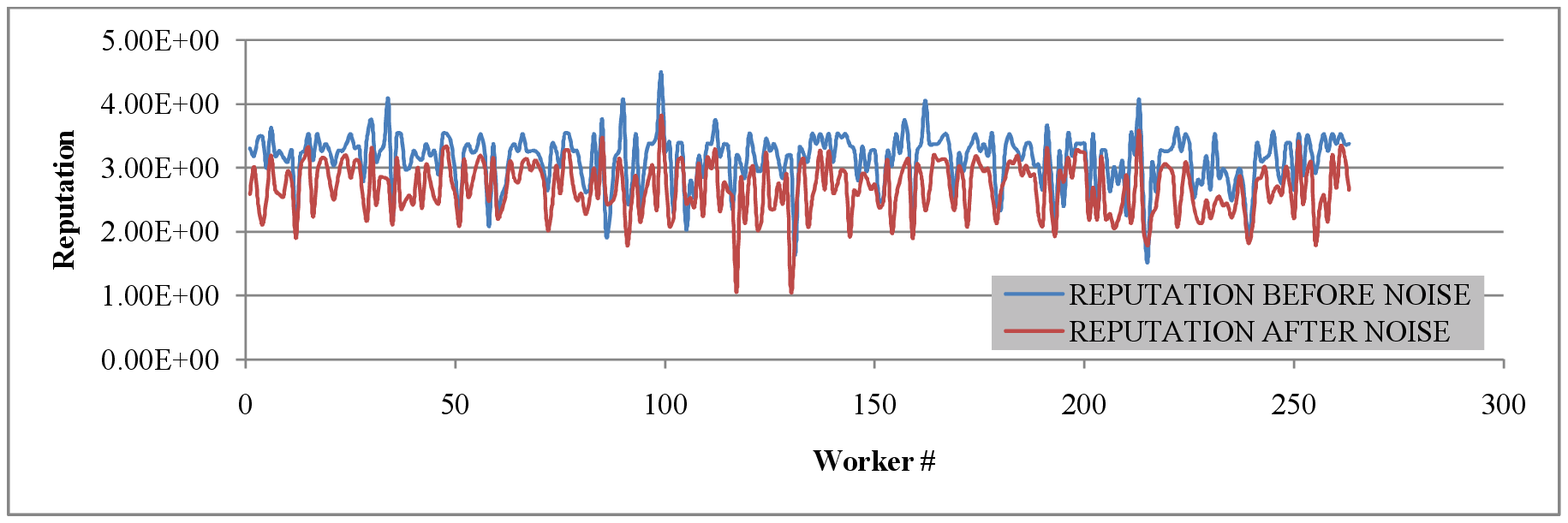}
\label{fig:prd}
}
\label{fig:difsc}
\caption[]{Reputations ranks before and after adding noise to dataset in three models}
\end{figure}

\begin{table*}
  \begin{center}
    \rowcolors{1}{}{lightgray}
    \begin{tabular}{| p{1.0cm} | p{1.5cm} | p{1.8cm} | p{1.8cm} | p{1.5cm} |p{1.5cm}|}
    \hline
    \textbf{User ID} & \textbf{User Name} &\multicolumn{2}{c|}{\textbf{Our Model}}& \textbf{ebay \newline Reputa-tion} & \textbf{Pagerank Reputation} \\
    \cline{3-4}
    & & \textbf{Reputation} & \textbf{Weight} &&\\
    \hline
    1 & taoster & 0.25 & 2.0 & 2.0 & 3.0 \\
    \hline
    2 & anthony & 1.0 & 0.1250 & 1.0 & 1.5 \\
    \hline
    .... & .... & .... & .... & .... & .... \\
    \hline
    948 & borisblue & 2.96 & 1.032 & 2.66 & 2.99 \\
    \hline
    .... & .... & .... & .... & .... & .... \\
    \hline
    1038 & elonka & 2.91 & 19.123 & 2.76 & 2.92 \\
    \hline
    .... & .... & .... & .... & .... & .... \\
    \hline
    \end{tabular}
  \end{center}
  \caption{Samples of Calculated Reputations.}
  \label{reps}
\end{table*}

As shown in figure \ref{difs}, 82.5 percent of the workers in our model experienced changes less than 10\% in their reputation; the corresponding fraction of users whose reputation rank changed less than 10\% is only 63.1\% for eBay model and 41.8\% for the page rank model. So, in comparison with eBay and pagerank models, our model is more robust against manipulations in reputation by unfair evaluations.

Figures \ref{fig:om} to \ref{fig:pr} show the distribution of changes in the reputation of the workers in these three models. We note that there are some users whose reputations in our model have not changed but in others have. To better compare the changes, we have chosen just 262 persons whose reputations in our model has changed and compared it with other models. As shown in figure \ref{fig:om} the changes in our model are distributed with an average of 0.167 and a standard deviation of 0.21. The average and the standard deviation for eBay model are 0.261 and 0.061 (figure \ref{fig:ebay}) and 0.464 and 0.319 for pagerank, respectively (figure \ref{fig:pr}). This implies that changes in the reputation of the workers in our model are mostly in range of $[0, 0.377]$ (the average change of 0.167 plus one standard deviation of 0.21). For ebay most of the changes fall in the range of $[2.0, 0.322]$ (average of 0.261) and finally for the pagerank in the range of $[0.145, 0.783]$ (average of 0.464). This shows that the average change in reputation of the workers due to such attack is in our model significantly lower than in the other two; thus, our model is more robust against unfair evaluations.

The Figure \ref{fig:difsc} also display a comparative view of changes happened in the reputation of the workers after inserting 20\% noise to the dataset. As it is obvious from the figure, most of the workers has reputation ranks of the value more than 2, thus most of the workers have been attacked by noises. Figure \ref{fig:omd}, shows the reputations calculated based on our model, Figure \ref{fig:ebayd} is related to ebay model and Figure \ref{fig:prd} is related to the pagerank model. The figures illustrate that the difference between the two reputation ranks calculated for users in our model is less than the other two models.

Our model also provides weights for calculated reputations. As shown in table \ref{reps}, the reputations calculated in ebay and pagerank models (and consequently all other similar models like Amazon and EigenTrust) are just one scalar value and it is very hard to judge the credibility of such calculated reputation, or to compare two reputation ranks just using their values. For our model the story is different. Every reputation comes with a corresponding weight showing the credibility of such reputation. This makes it easier and much more reliable to compare workers even when they have similar reputations. For example reputation of `borisblue' in table \ref{reps} is 2.96 and higher than reputation of the `elonka' which is 2.91. In terms of reputation values, `borisblue' is more trustworthy than `elonka' but by looking at the corresponding weights of reputations we realize that we can trust `elonka' more than `borisblue' because the weight of reputation of `borisblue', equal to 1.032, is more than eighteen times smaller than the weight of reputation of `elonka' which is 19.123. Thus, the reputation rank of `elonka' is far more credible than the reputation rank of `borisblue', making 'elonika' a preferred worker, despite its lower reputation. Thus, annotating reputation with a corresponding weight which indicates the credibility of such rank helps minimize the risk of choosing workers sub-optimally due to unreliable reputation ranks.

\begin{figure}
\centering
\includegraphics[scale=0.2]{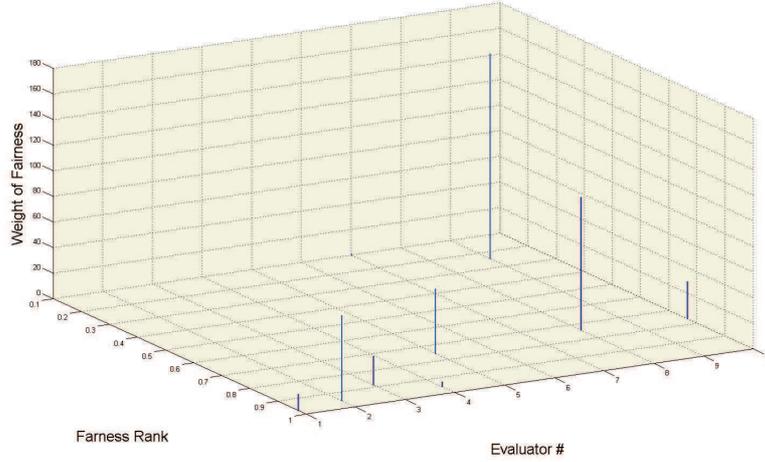}
\caption[My Image] {The schematic view of calculated Degrees of fairness}
\label{crds}
\end{figure}

The descriptive nature of reputation ranks illustrated in Table \ref{reps} is also applicable to the calculated fairness ranks. The Figure \ref{crds} schematically shows how fairness ranks of the evaluators can be compared in a 3D comparison chart. The fairness ranks are characterized by their value, their corresponding wight and the owner of the rank. Only 10 evaluators are chosen to display in the Figure \ref{crds} because showing all evaluators will decreases the readability of the chart while does not increase any added value. By Figure \ref{crds} we just intend to show ho informative are fairness ranks while are supported by wights in comparison with just presenting them as ordinal values.

In Figure \ref{crds}, $R_7$ has the degree of fairness of 0.1 with the weight of 1.0. It mean that although  $R_7$ has a low fairness rank but we can not be sure that she is that unfair, because we do not have enough evidences for that. $R_9$ has fairness rank of $0.9$ with the weight of $160.019$, thus we are sure that she is not a fair evaluator. On the other hand, $R_8$ has fairness rank of$0.742$ with the weight of $103.77$, so she is almost a fair evaluator. The $R_6$ has fairness rank of value $1.0$ which is the highest possible rank, but we do not have enough supporting evidences to argue that she is a fair evaluator since the weight of her fairness rank is $0.177$.

\section{Related Work}
There are two categories of studies related to our research.
\subsection{Crowd enhanced platforms}
The Amazon Mechanical Turk is a general-purpose online marketplace suitable for doing simple crowdsourcing tasks called Human Intelligence Task (HIT). There is no any metric called reputation in MTurk but there are some other metrics showing the trustworthiness of the workers like percentage of accepted or submitted HITS. There are no mechanisms in MTurk for detecting unfair evaluations and workers are highly vulnerable against misbehavior.

eBay is another crowd enhanced system in which the people sell and buy goods. People evaluate each other when they involve in transactions and based on these feedbacks, a reputation is built for the person as a seller or buyer or both. As we shown here, ebay reputation model is also  vulnerable against unfair evaluations.

StackOverflow\footnote{http://stackoverflow.com} is a question and answering web site. Users in StackOverflow can ask their questions, answer to questions asked by others and vote on the quality of the questions asked or answers provided by other community members. regarding received votes, a reputation is calculated for every member. There are no means for identifying unfair contributions in StackOverflow and users can easily manipulate reputations calculated for workers.

\subsection{Research Tools and Prototypes}
Noor and Sheng \cite{shengtrust} have proposed a trust management framework for cloud environments but it is very similar to reputation concept in crowdsourcing era and their idea is close to our work. They propose a credibility model for identifying unfair evaluations by using the concept of majority. They calculate an experience degree for every consumer evaluating services and apply it to the aggregation of his votes to eliminate votes form dishonest evaluators. The problem is that people sometimes are fair with most of the people while they are unfair just with a few number of people. In this case the unfairness of the evaluator will not be detected due to large number of fair votes. The other problem with Noor et.al model is that time and credit are not considered in calculation of trust and also experience of the customer.

Pagerank\cite{pagerank} is one of the most popular reputation management algorithms which employs the reputation of the evaluator in calculating the reputation of a worker. The votes given by highly reputable people are more important than votes of low reputable workers in pagerank model. This model is used by Google to rank pages in the internet. As we have shown in this work, pagerank does not employ any means for identifying unfair evaluations and is weak against unfair evaluation attacks. It also does not consider time in the reputation calculations.

EigenTrust\cite{eigentrust} is a popular trust model which is built based on the pagerank algorithm and tried to solve the problem of unfair evaluations. EigenTrust supposes that there are some pre-trusted users in the system in which we can trust and it is evident that this assumption is not applicable to most of the existing crowdsourcing systems like question and answering systems or online marketplaces. It is also proven that EigenTrust is not robust against unfair evaluations\cite{EigenCollusion}.

\section{Discussion and Conclusion}
In this paper we have proposed a model for reputation management in crowdsourcing environments. We have introduced an analytic model for calculating a more accurate, realistic and reliable reputation rank of workers by taking into account the time, the credit amount and more importantly the credibility of the evaluators. We have also proposed a model for calculating a reputation for assessing how fair an evaluator in evaluating the contributions of the workers. We use this degree of fairness to distinguish the honest evaluators from dishonest ones who cast unfair votes. We have validated our model using experimental evaluations and compared the accuracy and credibility of our model with eBay and pagerank models. The results presented show that our model is more robust against manipulating the reputation of workers by unfair evaluations than two commonly used methods (eBay and Pagerank).

In the real world workers are involved in many crowdsourcing tasks, and our method very effectively utilizes this fact to make it harder to manipulate worker's reputation by dishonest and unfair evaluations. The more activities the worker has, the more evaluations are needed to create a major change in her reputation. So, the experienced users that have lots of activities will benefit from more robust and realistic reputation scores.

For the novice workers or workers with few activities, because of the small number of evaluations that build up their reputation scores, it is easier to change the worker's reputation by unfair evaluations; however, when the overall number of the  activities of the  user increases in time, those unfair evaluations will be detected and gradually the reputation of the worker will be corrected and unfair evaluations will be essentially ignored by our method for calculating reputation of workers.

As future work, we plan to extend our model to identify colluding groups and protect workers against badmouthing and ballot stuffing. We are in the process of building a flexible people evaluation tool based on our model which can be seamlessly integrated with the existing crowdsourcing platforms.

%
\bibliographystyle{abbrv}
\bibliography{KDD_Tech_Rep.bbl}
\end{document}